\documentclass[preprint,aps,superscriptaddress,floatfix, preprintnumbers,amsmath,amssymb,nofootinbib]{revtex4}
\usepackage{graphicx}
\usepackage{dcolumn}
\usepackage{bm}

\newcommand{\be}{\begin{equation}}
\newcommand{\ee}{\end{equation}}
\newcommand{\bea}{\begin{eqnarray}}
\newcommand{\eea}{\end{eqnarray}}

\begin{document}
\title{Nanomachines driven by thermal bath}
\author{Zhiping Xu}
\affiliation{
Department of Engineering Mechanics, Tsinghua University, Beijing
100084, China
}
\author{Quanshui Zheng}
\affiliation{
Department of Engineering Mechanics, Tsinghua University, Beijing
100084, China
}
\date{\today}

\begin{abstract}
Multiwalled carbon nanotubes based nanomachines driven by thermal bath are proposed in this Letter. Atomistic simulation shows that intershell translational and rotational motion can be activated at sufficiently high temperature, where the van der Waals potential barrier is accessible by the thermal fluctuation. The thermal fluctuation is further identified to be consistent with the equipartition theorem. This model can be used to construct one dimensional devices driven by thermal bath or nano-channels for molecule transport.
\end{abstract}

\maketitle

In 1828 Robert Brown reported his observation on the rapid and irregular motion of pollens suspended in water \cite{Brown-1828}. The phenomenon was explained theoretically later by the combinational effects of a fluctuating force and a viscous drag force resulting from collisions with solvent molecules \cite{Einstein-1956}. Both the phenomena and the theory are immeasurably important in modern physics and continual development has been made for centuries \cite{Parisi-2005}. With the advent of recently developed nanotechnology and inspiration from the biological world, people now are interested in constructing small devices working at molecular level and activated by thermal noise. One example is the so-called Brownian ratchet \cite{Astumian-1997}, where particles drift unidirectionally over a switchable asymmetric and periodic potential profile. In this Letter we present multiwalled carbon nanotubes based nanomachines, which can be fuled by thermal fluctuation. With appropriate control, it can act as one-dimensional device working at a length and time scale of nano.

Carbon nanotubes, composed by one or more layers of cylindrical graphene shells, attracted much interest from various disciplines recently \cite{Baughman-2002}. These intriguing quasi-one dimensional structures possess outstanding material properties, such as high aspect ratio, chemical and thermal stability, high modulus and strength, and good transport properties. It also holds the promise for the building blocks in the next-generation nano-electromechanical systems (NEMs) \cite{Craighead-2000}. Except for their unique mechanical and electronic properties, one of the most fascinating features of this novel material is that within the graphene shells, the sp$^{2}$ bonds between carbon atoms are the strongest bond in nature, while between different shells only very weak van der Waals interactions. This high anisotropy provides the carbon nanotubes based devices with great intershell mobilities without losing structure stability as operating.

Especially, model devices, such as oscillators and actuators based on multiwalled carbon nanotubes, were proposed \cite{Zheng-2002} and experimentally implemented \cite{Fennimore-2003} in past few years. These devices, where graphene shells act as moving parts, have many impressive properties such as ultra high frequency and atomic precision. In the axial oscillator model \cite{Cumings-2000, Zheng-2002}, the inner core within a open-end holder was pulled out at first and released then. The intershell restoring van der Waals force would drag the core back into the holder. Because of its inertia and the negligible resistance force \cite{Cumings-2000, Yu-2000}, the core will slide back and forth afterwards. The operating frequency was predicted to be up to gigahertz (GHz). Because of its great potential in engineering applications, lots of studies have been carried out on the dynamics, mainly through atomistic simulations \cite{Guo-2003, Servantie-2003, Tangney-2004, Zhao-2003, Zhao-2006}. These works confirmed the high frequency on the order of GHz, but the energy dissipation was found to be rather severely. The axial oscillation initialized by extrusion decayed quickly in few nanoseconds. It's far away from our intuition and can not be explained by the previously predicted small static friction force.

In fact, in such a device working at high speed on the order of 1.0 nm/ps, sliding will dissipate by leaking kinetic energy to other internal modes such as bending and radial breathing \cite{ Zhao-2003, Zhao-2006, Servantie-2006}. The rate of leaking was found to depend on the temperature, chiralities and the system size. More importantly, open tube ends in the finite systems also make significant contribution \cite{Tangney-2004}. Thus from an application point of view, low dissipation can be expected by lowering commensurabilities and avoiding the ends crossing of nanotubes during operating.

\begin{figure}
\begin{center}
\includegraphics[scale=0.35]{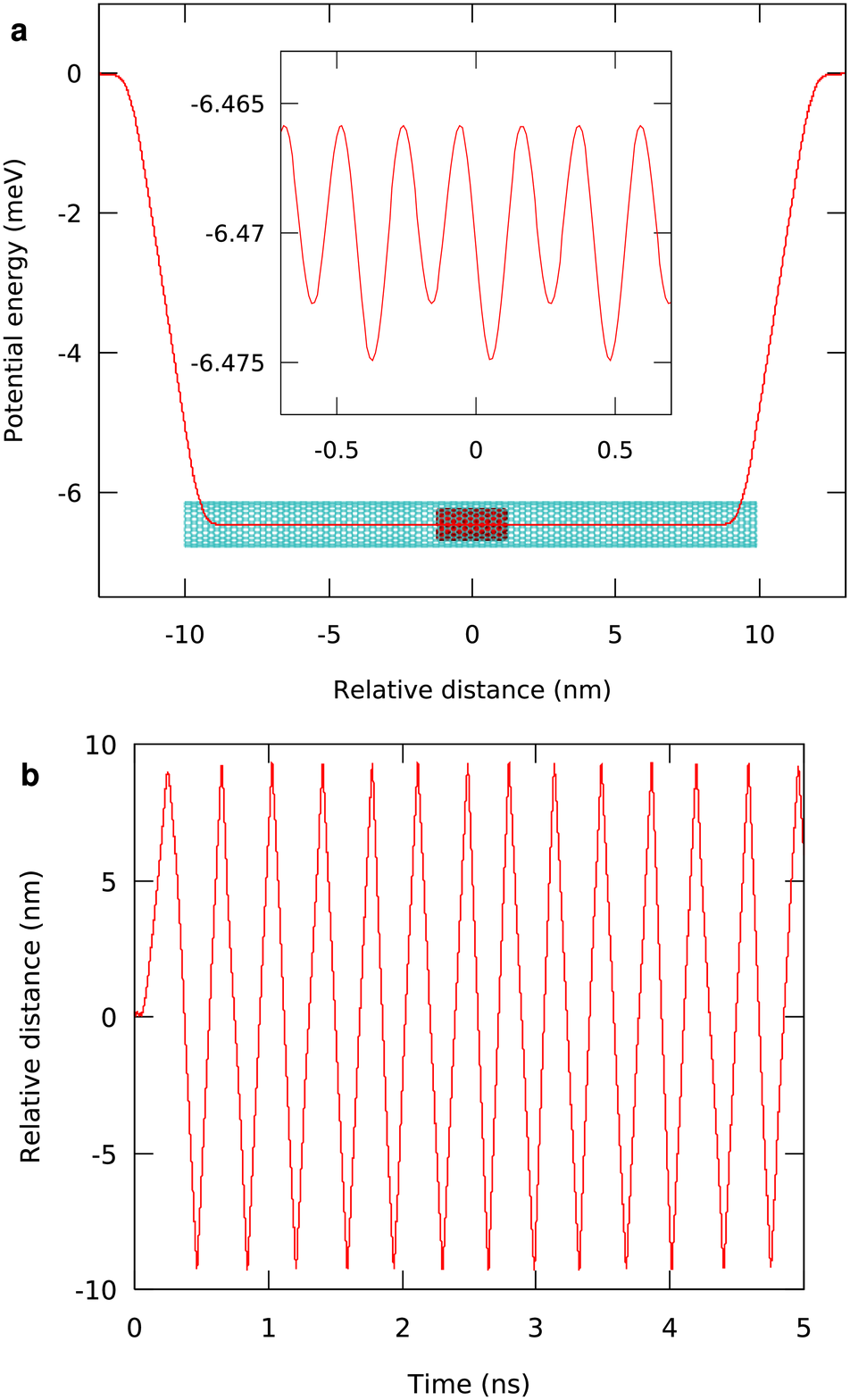}
\end{center}
\caption{
(Color online) (a) Atomic structure and axial potential energy profile of the DWNTs (7,7)@(21,0), with a typical intershell distance of 0.34 nm. The inner core (red), with a length of 2.5 nm, is confined by the ends repulsion potential well $V_{\mathrm{ends}} = 6.47$ meV per atom. While inside the well, there are three orders smaller corrugation $V_{\mathrm{loc}} = 0.009$ meV per atom provided by the interaction with the 20 nm long outer tube (gray), as shown in the inset. (b) Axial sliding of the DWNTs, initialized by an axial speed at 100K. The oscillation amplitude does not decay during the 5 nanoseconds simulation, which suggests ultra-low dissipation of the system.
}
\label{figure1}
\end{figure} 

To verify this, we study the incommensurate double-walled carbon nanotubes (DWNTs) (7,7)@(21,0) as shown in Figure 1(a) by molecular dynamics simulation \cite{Gromacs}. The Dreiding forcefield \cite{Guo-1991} is used to describe the interatomic interaction. Firstly the intertube van der Waals potential energy profile as a function of the axial displacement is illustrated in Figure 1(a). It can be seen that the core is confined within a steep well of $V_{\mathrm{ends}} = 6.47$ meV per atom provided by the tube ends repulsion. However inside the well, the local corrugation against sliding is much smaller, about $V_{\mathrm{loc}} = 0.009$ meV per atom, which is comparable to the thermal energy at room temperature. So the dissipation of sliding inside the ends potential energy wells should be small by avoiding the ends effect following the arguments on the dissipation mechanism given above. In our simulation the system is firstly coupled with a $100$ K bath for 200 ps. After the thermal equilibria, an axial velocity is assigned to the inner tube and the nanotubes start to oscillate between the outer tube ends. The relative displacement evolution of the DWNTs is shown in Figure 1(b). The axial oscillation is found to have negligible dissipation during the 5 nanosecond simulation. This result contradicts those reported \cite{Servantie-2003, Zhao-2003, Tangney-2004}, in which the length of the DWNTs are comparable and the inner tube is pulled out of the holder, so that tube ends cross and instabilities are involved \cite{Tangney-2004}. The ultra-low dissipation observed here provides possibilities for energy supplement and experimental observation with a much larger timescale than nanoseconds.

Physically speaking, the dissipation of sliding is the redistributing process from the sliding to internal modes. When the system reaches equilibrium, each mode, including axial sliding, holds a portion of the kinetic energy $\frac{1}{2}k_{\mathrm{B}}T$ following the equipartition theorem. And the DWNTs will oscillate, in a fluctuative manner, relative to each other. This kind of fluctuative oscillation was observed by Servantie and Gaspard recently \cite{Servantie-2006}. However, as mentioned above, because the ends effects are involved in their work, the local potential wells at the equilibrium state are too steep, so the fluctuative the position $\langle r^{2} \rangle^{\frac{1}{2}} = \sqrt{k_{\mathrm{B}}T/k}$ is bounded in few Angstroms, where $k$ is the effective elastic constant of the well. In contrast, the axial barrier in our model is negligibly small here, so large amplitude diffusion of the inner core could be expected. Considering the intershell sliding motion as a normal mode, and the reduced mass as $\mu = \frac{m_{\mathrm{in}}m_{\mathrm{out}}}{m_{\mathrm{in}}+m_{\mathrm{out}}}$, where $m_{\mathrm{in}}$ and $m_{\mathrm{out}}$ are the mass of the inner and outer tube respectively. The velocity needed to overcome the local barrier $V_{\mathrm{loc}}$ is estimated to be $v_{\mathrm{loc}} \sim 0.05$ nm/ps. As the axial speed excited by the thermal noise exceeds this critical value, diffusion over the local barriers will be established.

\begin{figure}
\begin{center}
\includegraphics[scale=0.5]{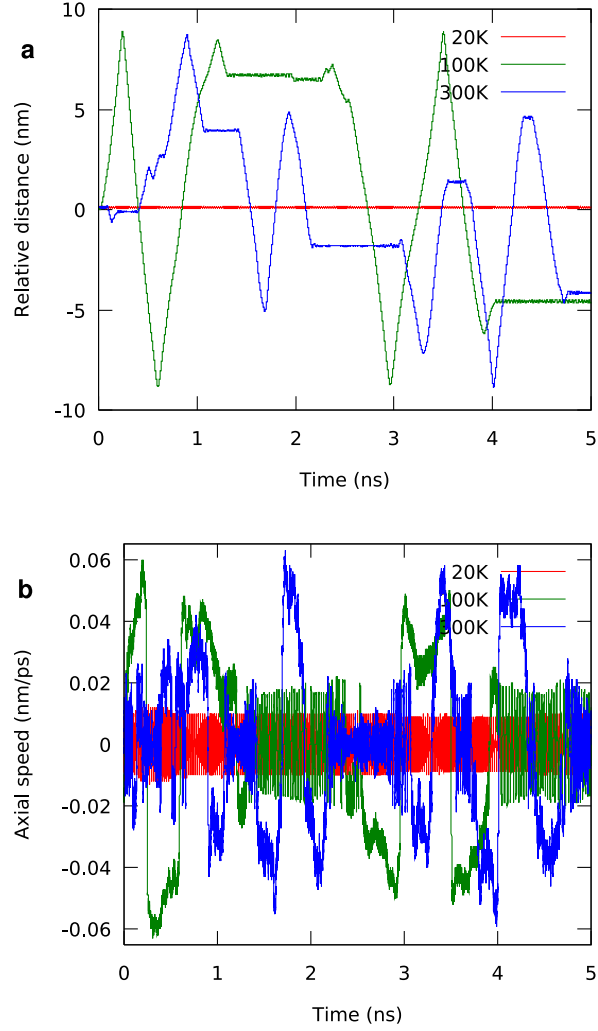}
\end{center}
\caption{
(Color online) (a) Axial displacement of the DWNTs coupling with a thermal bath at $20$ K (red), $100$ K (green), and $300$ K (blue) respectively. The fluctuative oscillation at low temperature is confined in the local potential barrier of 0.2nm as shown in Figure 1 (a), while in temperature higher than $100$ K large amplitude axial sliding is excited. Besides the constant speed sliding over local barriers, there are also some flat regions in the displacement curve. In these regions the axial diffusion is temporarily trapped, resulting from the thermally induced structure distortion as explained in the text later. (b) Corresponding axial velocity profile. There exists a critical value of the sliding speed $v_{\mathrm{cr}}$, values higher than that correspond to the linear diffusion in (a), and lower values correspond to the trapped region. The critical value $v_{\mathrm{cr}} = 0.05$ nm/ps, read from the curve, coinciding well to the estimated velocity $v_{\mathrm{loc}}$ to overcome the local potential barrier.
}
\label{figure2}
\end{figure} 

For this purpose, the DWNTs are investigated under thermal coupling. In the simulations, the outer tube is connected to Nos\'e-Hoover thermostats at various temperature from $20$ K to $300$ K, and the core feels the thermal coupling through collision with the outer tube only. After equilibrating time of $200$ ps, the data is acquired as integrating the equations of motion and plotted in Figure 2. At lower temperature, $20$ K for example, the sliding motion is trapped within the $0.2$ nm wide local barriers $V_{\mathrm{loc}}$. The axial speed fluctuates with a peak value of $0.01$ nm/ps which is much less than $v_{\mathrm{loc}}$. However at higher temperature as $100$ K and $300$ K, where more kinetic energy distributed on the sliding mode, global diffusion and oscillation between the tube ends is excited. One remarkable feature of this thermally excited oscillation is its large amplitude extending to the whole outer tube and high frequency on the order of GHz, which can be further tuned by the design of DWNTs.

The oscillation behavior observed can be resolved into two regimes: (1) the bounded fluctuation state (BFS) (2) large-scale diffusion extending to the whole outer tube. In the BFS, axial speed fluctuates but is lower than the critical value $v_{\mathrm{loc}}$. Specifically, for the thermally excited oscillation at $100$ K, kinetic energy distributed on sliding from the equipartition theorem is $\frac{1}{2}k_{\mathrm{B}}T$, corresponding to average axial speed of $0.016$ nm/ps, which is consistent with the simulation results. This consistency suggests the equilibrium fluctuation nature of the BFS. When the axial speed exceeds $v_{\mathrm{loc}}$ at some moment, the inner tube jumps out of the local barriers and diffuse over them. Because of the ultra-low sliding dissipation of the DWNTs, the diffusion speed almost keeps constant at a value close to $v_{\mathrm{loc}}$. 

In addition, arguments should be addressed that the sliding fueled by the thermal bath observed in the DWNTs is not a violation of the second law in thermodynamics. The oscillational behavior is fluctuative and the axial speeds, the trapped time and the thus the oscillation frequency are not deterministic. For application, external controls such as external field or temperature gradient \cite{Ouyang-2004}, which introducing energy, must be applied to obtain a regular and controllable motion.

Besides of the axial diffusion, angular motion of the tubes is also observed to be excited in the simulations. In contrast to the axial oscillation, unidirectional rotation is driven with average speed at $0.04$ rad/ps. However, the amplitude is also fluctuative as shown in Figure 3.

\begin{figure}[bottom]
\begin{center}
\includegraphics[scale=0.5]{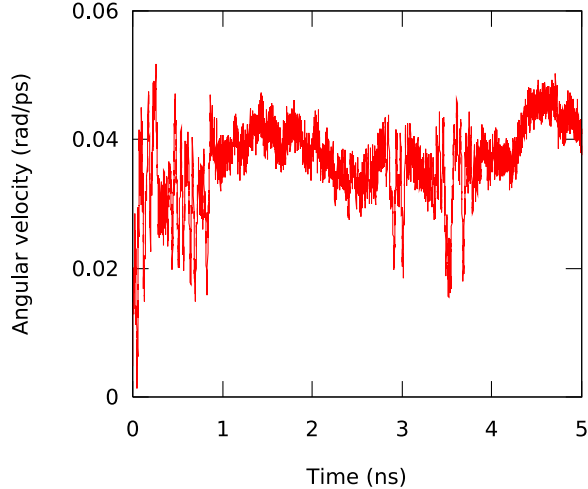}
\end{center}
\caption{
Relative angular velocity of the double-walled nanotube, coupled with thermal bath at $100$ K. The velocity fluctuates with a mean value of $0.04$ rad/ps.
}
\label{figure3}
\end{figure} 

Finally we consider the irregular parts of the oscillation profile shown in Figure 2(a), especially the flat regions in the curve of sliding amplitude at $100$ K. The atomic structure of the nanotubes at the corresponding simulation time is extracted and shown in Figure 4. Structure distortion, which causes the outer tube to deviate from its cylindrical geometry, has been found. The local potential profile against sliding is calculated and plotted. It is found that the tube deformation induces an enhanced local barrier of $0.025$ meV per atom, which requires a larger $v_{\mathrm{cr}} = 0.083$ nm/ps. This value can not be exceed by the fluctuative speed shown in Figure 2. Thus the linear diffusion of the inner tube is trapped again and will not continue to slide until the local distortion is released. The distortion of graphene lattice results from the equilibrium fluctuation of other internal modes such as the bending and rocking modes \cite{Zhao-2003}. Using multiwalled nanotubes or apply constraints on the outer tube may help to inhibit the deformation and avoid these instabilities.

We have also performed simulations under higher temperature $1000$ K and $1500$ K. The distortion of the graphene shells are much more significant, and the global axial motion is frequently inhibited within few nanometers, although higher axial speeds over $0.1$ nm/ps are excited.

\begin{figure}
\begin{center}
\includegraphics[scale=0.50]{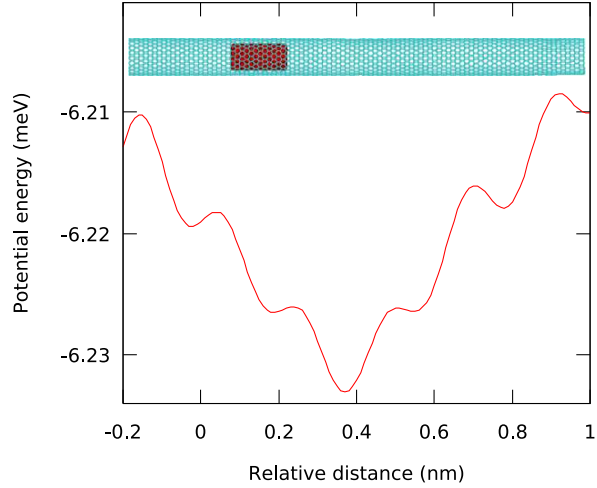}
\end{center}
\caption{
(Color online) The trapped state at $100$ K corresponding to the flat region in displacement curve, at $t = $ 4.8 ns, plotted in Figure 2(a). The cylindrical lattice of outer tube is found to be distorted. This imperfection causes an enhanced local well of  $0.025$ meV per atom against sliding, which requires a larger axial speed ( $0.083$ nm/ps) to overcome.
}
\label{figure4}
\end{figure} 

In conclusion, DWNTs based nanomachines driven by thermal bath are proposed in this Letter. Coupling with thermal bath at certain temperature, the inner tube can oscillate within the ends repulsion potential wells. The oscillation is a fluctuative phenomenon, with amplitude and operating frequency decided by the length of the outer tube. With external control applied, the nanomachine can be used to fabricate artificial nano devices. The physics behind this novel phenomenon is the kinetic energy hold by the translation or rotation at the thermal equilibrium is sufficient to overcome the corresponding potential barrier provided by the atomistically smooth potential surface of the graphene shell. Furthermore the ends potential provides a working space within which the severe dissipation by ends crossing is avoided. We have also investigated the confined dynamics of C$_{60}$ inside carbon nanotubes. Similar fluctuative axial and angular motion are excited, furthermore the the potential profile can be tuned by varying the diameter and chirality of nanotube. Thus our observation suggests also the possibilities of fast molecule transport channel assisted by thermal noise and tuned by the tube geometry. On the other hand, this model system may also be theoretically interesting for studies of the nanoscale thermodynamics \cite{Bustamante-2005}. However, there still remain great but exciting experimental challenges for observing and controlling the nanoscale dynamics.

\begin{acknowledgments}
This work is supported by NSFC through Grants 10332020, 10252001 and 50518003.
\end{acknowledgments}


\begin{thebibliography}{}
\bibitem{Brown-1828}
R. Brown, \textit{Philos. Mag.} \textbf{4}, 161 (1828).
\bibitem{Einstein-1956}
A. Einstein, R. F\"urth, A. D. Cowper, \textit{Investigations on the theory of the Brownian movement}, New York: Dover, 1956.
\bibitem{Parisi-2005}
G. Parisi, \textit{Nature} \textbf{433}, 221 (2005).
\bibitem{Astumian-1997}
R. D. Astumian, \textit{Science} \textbf{276}, 917 (1997).
\bibitem{Baughman-2002}
R. H. Baughman, A. A. Zakhldov and W. A. de Heer, \textit{Science} \textbf{297}, 787 (2002).
\bibitem{Craighead-2000}
H. G. Craighead, \textit{Science} \textbf{290}, 1532 (2000).
\bibitem{Zheng-2002}
Q. Zheng and Q. Jiang, \textit{Phys. Rev. Lett.} \textbf{90}, 045503 (2002).
\bibitem{Fennimore-2003}
A. M. Fennimore \textit{et al.}, \textit{Nature} \textbf{424}, 408 (2003).
\bibitem{Cumings-2000}
J. Cumings and A. Zettl, \textit{Science} \textbf{289}, 602 (2000).
\bibitem{Yu-2000}
M. F. Yu \textit{et al.}, \textit{Science} \textbf{287}, 637 2000; M. F. Yu, B. I. Yakobson and R. S. Ruoff, \textit{J. Phys. Chem. B} \textbf{104}, 8764 (2000).
\bibitem{Guo-2003}
W. Guo \textit{et al.}, \textit{Phys. Rev. Lett.} \textbf{91} 125501(2003).
\bibitem{Servantie-2003}
J. Servantie and P. Gaspard, \textit{Phys. Rev. Lett.} \textbf{91}, 185503 (2003).
\bibitem{Tangney-2004}
P. Tangney, S. G. Louie and M. L. Cohen, \textit{Phys. Rev. Lett.} \textbf{93}, 065503 (2004).
\bibitem{Zhao-2003}
Y. Zhao \textit{et al.}, \textit{Phys. Rev. Lett.} \textbf{91}, 175504 (2003).
\bibitem{Zhao-2006}
Y. Zhao, C. C. Ma, L. Wong, G. Chen, Z. Xu, Q. Zheng, Q. Jiang and A. T. Chwang, \textit{Nanotechnology} \textbf{17}, 1032 (2006).
\bibitem{Servantie-2006}
J. Servantie and P. Gaspard, \textit{Phys. Rev. B} \textbf{73}, 125428 (2006).
\bibitem{Guo-1991}
Y. Guo, N. Karasawa and W. A. Goddard, \textit{Nature} (London) \textbf{351}, 464 (1991).
\bibitem{Gromacs}
E. Lindahl, B. Hess and D. van der Spoel, \textit{J. Mol. Mod.} \textbf{7}, 306 (2001); H. J. C. Berendsen, D. van der Spoel and R. van Drunen, \textit{Comp. Phys. Comm.} \textbf{91}, 43 (1995).
\bibitem{Ouyang-2004}
Z. C. Tu and Z. C. Ou-Yang, \textit{J. Phys.: Condens. Matter} \textbf{16}, 1287 (2004); Z. C. Tu and X. Hu, \textit{Phys. Rev. B} \textbf{72}, 033404 (2005).
\bibitem{Bustamante-2005}
C. Bustamante, J. Liphardt and F. Ritort, \textit{Physics Today} 43, July (2005).
\end{thebibliography}
\end{document}